\documentclass[preprint]{ore}
\usetikzlibrary{arrows.meta,positioning,calc}
\usepackage{subcaption}
\usepackage[capitalise,noabbrev]{cleveref}
\newenvironment{oretablec}[1]
  {\par\centering\begin{tabular}{#1}\hline}
  {\end{tabular}\par}

\articletype{Research Article}  

\title[Automated classification of modal displacements]{Automated classification of modal displacements}

\author*[1]{Gabriele Dessena}[0000-0001-7394-9303]
\author[2]{Marco Civera}[0000-0003-0414-7440]
\author[1]{Saray Undiano Echániz}
\author[3]{Oscar Bonilla-Manrique}[0000-0003-0541-8310]

\email{gdessena@ing.uc3m.es}

\affil[1]{Department of Aerospace Engineering, Universidad Carlos III de Madrid, Avenida de la Universidad 30, 28911 Leganés, Spain}
\affil[2]{Department of Structural, Geotechnical, and Building Engineering, Politecnico di Torino, Corso Duca degli Abruzzi 24, 10129 Turin, Italy}
\affil[3]{Electronic Technology Department, Universidad Carlos III de Madrid, Avenida de la Universidad 30, 28911 Leganés, Spain}

\shortauthors{Dessena G, Civera M, Undiano Echániz S, Bonilla-Manrique O}

\articleversion{1}
\peerreviewstatus{awaiting peer review}

\preprintlabel{Preprint}               


\authorroles{\textbf{Dessena G}: Methodology, Software, Validation, Formal Analysis, Investigation, Resources, Data Curation, Writing -- Original Draft, Writing -- Review \& Editing, Visualization, Supervision, Project administration, Funding acquisition; \textbf{Civera M}: Methodology, Validation, Resources, Writing - Review \& Editing, Visualization, Supervision; \textbf{Undiano Echániz S}: Conceptualisation, Software, Investigation; \textbf{Bonilla-Manrique O}: Resources, Writing - Review \& Editing, Funding acquisition.}  
\competinginterests{No competing interests to disclose.}
\grantinformation{The first author is supported by grant JDC2024-055593-I funded by MICIU/AEI /10.13039/501100011033 and by ESF+. This work is supported by the Madrid Government (\emph{Comunidad de Madrid} – Spain) under the Multiannual Agreement with the Universidad Carlos III de Madrid (\href{https://researchportal.uc3m.es/display/act564873}{IA\_aCTRl-CM-UC3M}).}

\begin{document}

\maketitle

\begin{abstract}
\abstractsection{Background}
The bending, torsional, or coupled character of a mode shape helps determine the dynamic behaviour of aerospace and civil structures. 
In vibration testing, it is assigned by inspecting mode shape plots, and depends on the analyst.

\abstractsection{Methods}
This work proposes four indices to describe the motion of a mode from a few sensor vectors: a bending score, a torsion score, a twist share and an in-plane ratio. They are assessed on the first eight finite element modes of a tapered plate with a swept trailing edge. They are applied to identified XB-2 wing modes and to the published modes of the Sheraton Universal Hotel, a reinforced concrete (RC) building.

\abstractsection{Results}
On the plate, the twist share separates bending, coupled and torsion modes. The XB-2 modes include one bending and three coupled modes (one dominated by in-plane bending and two by out-of-plane bending). The coupled mode dominated by in-plane bending (15.25 Hz, in-plane ratio 1.33) was not investigated in earlier campaigns on the wing. On the RC building, all five modes are bending (twist share below 0.0005).

\abstractsection{Conclusions}
The twist share and the in-plane ratio indicate the type of each mode. Hence, modes are classified without visually inspecting the mode shapes. Code and processed data are openly released.
\end{abstract}

\begin{plainsummary}
When a structure such as an aircraft wing or a building vibrates, each of its natural vibration patterns bends it, twists it, or does both. 
Structural engineers usually classify these vibration patterns by inspecting plots of the corresponding mode shapes.
This work replaces that visual step with four indices, computed from the identified motion at a few accelerometers. These state how much a vibration pattern bends, twists or moves within the main plane of the structure. The numbers are checked on a computer model of a tapered plate and on a real flexible wing, for aeronautical applications. They are also applied to a high-rise reinforced concrete building to examine their use on a civil engineering structure.
\end{plainsummary}


\makeinfobox

\clearpage
\newgeometry{top=2.3cm, bottom=2.2cm, left=3.1cm, right=3.1cm, headheight=26pt, headsep=12pt, footskip=1.0cm}
\fancyhfoffset[L,R]{1.1cm}

\section{Introduction}
Modal parameters, namely the natural frequencies ($f_n$), the damping ratios ($\zeta_n$) and the mode shapes ($\boldsymbol{\phi}_n$), describe the linear dynamic behaviour of a structure and are the basis of experimental structural dynamics.\cite{Ewins2000} They are used for structural health monitoring (SHM),\cite{Fan2011} for finite element (FE) model updating\cite{Scalia2026} and for the aeroelastic assessment of aircraft.\cite{Dessena2025} Among them, $\boldsymbol{\phi}_n$ carry the spatial information of the response, that is, the relative amplitude and direction of the motion at each measured point.

In SHM, this spatial information is used to locate damage, and the methods based on the curvature of $\boldsymbol{\phi}_n$ are in general very effective at localising it.\cite{Fan2011} When few accelerometers are available, the measured $\boldsymbol{\phi}_n$ can be expanded before the curvature is computed, allowing damage to be located and quantified in steel and concrete beams \cite{Modesti2024}.
The wavelet transform of $\boldsymbol{\phi}_n$ also localises and quantifies damage in numerical beam models.\cite{Vila2024} In addition, modal correlation criteria built on $\boldsymbol{\phi}_n$ form the objective function of FE model-based damage identification from incomplete and noisy modal data.\cite{Georgioudakis2018} Similar criteria, which combine the healthy and damaged $\boldsymbol{\phi}_n$ with the change in $f_n$, detect damage and assess its severity in the ground vibration test (GVT) data of a full BAE Systems Hawk T1A airframe.\cite{Dessena2025a} In bridge monitoring, $\boldsymbol{\phi}_n$ are also reported to be less affected by environmental variability than $f_n$, and transverse mode shapes respond to scour.\cite{BukaVaivade2025} On rural bridges, low-cost accelerometers carried by vehicles detect deviations of the measured response from a learned baseline.\cite{Tantaroudas2026}

In FE model updating, the measured modal parameters help correct a model that disagrees with the test.\cite{Mottershead1993} The differences between measured and predicted $\boldsymbol{\phi}_n$ enter the residuals even when far fewer degrees of freedom are measured than modelled.\cite{Goge2003} This correction is preceded by the correlation of test and analysis,\cite{Ewins2000a} which pairs each measured mode with a predicted one through the Modal Assurance Criterion (MAC)\cite{Allemang1982} and its coordinate-wise  counterpart, the Coordinate Modal Assurance Criterion (COMAC).\cite{Lieven1988}
Pairing also tracks modes across successive identifications. This tracking is hampered by crossing modes\cite{Verboven2002} and, in flutter flight tests, it relies on a criterion that combines $\boldsymbol{\phi}_n$ and poles.\cite{Volkmar2026} In aeroelasticity, flutter analysis is built on the normal modes of the structure,\cite{Wright2014} taken from FE models matched to the $f_n$ and $\boldsymbol{\phi}_n$ measured in GVTs.\cite{Goge2003a}

The deformation pattern is used to classify the modes of a full aircraft as wing bending, wing torsion, or tail bending. \cite{Kerschen2013} 
Such labels are assigned by mapping the identified $\boldsymbol{\phi}_n$ onto the geometry and inspecting them, as on a PC-9A wing, where a wireframe overlay separates the first bending and first torsion modes.\cite{Das2026} This is critical to understanding how instability develops. For instance, in the very flexible wing of Pontillo et al.,\cite{Pontillo2025} the first in-plane mode is found to drive flutter.

For aircraft wings, the type of motion of each mode matters as much as its frequency, since the interaction between bending and torsion governs the onset of flexure--torsion flutter.\cite{Hancock1985} In GVT, the type of an identified mode is assigned by inspecting the plotted $\boldsymbol{\phi}_n$. This is the case for the flexible wing of the BEAm Reduction and Dynamic Scaling (BeaRDS) programme, the eXperimental BeaRDS 2 (XB-2), whose modes are labelled as bending or coupled from their vertical displacement.\cite{Dessena2022b} A further mode of this wing is not investigated in its GVT campaigns, since it is dominated by lagging.\cite{Dessena2022h} Nonetheless, the classification relies on the judgement of the analyst. It also becomes laborious when many modes, test configurations or monitoring sessions must be compared.

Torsion leaves a distinct signature only at sensors placed away from the twist axis. Hence, its detection from translational measurements is not trivial. 
Array-derived rotation has been used to estimate torsion from translational measurements and has been compared with direct rotational measurements.\cite{Gueguen2021} 
Torsional and bending modes are also studied separately, and the former are found more sensitive to damage.\cite{Aulakh2021} The MAC quantifies the similarity between two $\boldsymbol{\phi}_n$,\cite{Allemang1982} but it does not describe the type of motion of a single mode. Scalar indicators of a single identified mode, such as the consistent-mode indicator, rate the reliability of the mode rather than its type of motion.\cite{Pappa1993} Likewise, automated modal analysis selects the physical modes from a stabilisation diagram with thresholds set from the data, but it does not classify their motion.\cite{Reynders2012a}

Other methods resolve the motion of a mode into its components. For large FE models, the modes are named through their kinetic and strain energy fractions by direction and component, which require the mass and stiffness matrices of the model.\cite{Coppolino2011} For wind turbine blades, the modal amplitudes measured by three accelerometers at each cross-section are resolved into flapwise, edgewise and torsional deflections, and each mode is named after the dominant one in their spanwise curves.\cite{Larsen2002} The contribution of each component to a blade mode is also quantified from the displacements of the leading and trailing edges of each cross-section, and the sections are summed into the share of each component over the whole mode. This method is demonstrated on numerical and laser-vibrometer mode shapes of beam-type structures.\cite{Chen2022} On the other hand, graph convolutional networks trained on a dataset of FE mode shapes classify the modes of sport utility vehicle models.\cite{Tohmuang2025} However, these methods require a model, training data or the reading of spanwise curves, or they are formulated for the cross-sections of beam-type structures. In contrast, the indices proposed here describe the motion of a mode directly from a few sensor vectors, and they add two measures of the coherence of the motion and of its residual. In addition, each section is formed by any pair of sensors spaced transversely to the span axis, so that the indices also apply to a reinforced concrete (RC) building, with its height as the span axis.

This work proposes an automated classification of mode shapes from the three-dimensional motion vectors of a small set of sensors. 
In this context, ``automated'' means that the indices and the modal label are computed directly from the measured modal displacement vectors, without the visual inspection of mode-shape plots, using decision thresholds calibrated empirically on the cases presented here.
Four indices describe each mode. A bending score is computed from the sensor vectors, while a torsion score T is based on the directions of the residual and tangential vectors. 
Instead, a twist share is computed from the motion of pairs of sensors on opposite sides of the span axis, such as the leading and trailing edges of a wing. The fourth index is an in-plane ratio between the motion within the reference plane defined by the span direction and the transverse sensor-pair direction. 
Together, the four indices characterise the modal motion without requiring visual inspection of the mode shapes. A classification rule combines the twist share with the global and section-wise in-plane ratios, to assign labels describing the dominant and secondary motion.
The method is assessed on the FE modes of a tapered cantilever plate with a swept trailing edge, with the sensors reading on its top surface. The method is then applied to the mode shapes of the XB-2 wing identified through the real-valued improved Loewner Framework (iLF).\cite{Dessena2024} Finally, it is applied to the published mode shapes of a building, the Sheraton Universal Hotel, with its height taken as the span axis.

The contributions of this work are:
\begin{itemize}
  \item Definition of four indices that describe the motion of a mode from the sensor vectors, without the visual inspection of the mode shape;
  \item A rule that assigns labels describing the dominant and secondary motion of a mode from the twist share and the in-plane ratio, with bounds quantified from the three cases;
  \item Assessment of the indices on the FE modes of a tapered cantilever plate with a swept trailing edge, with the motion read at the top surface;
  \item Application of the indices to the mode shapes of the XB-2 flexible wing, identified from its GVT data;
  \item Application of the indices to the published mode shapes of a building, the Sheraton Universal Hotel, with its height taken as the span axis;
  \item Release of the code and of the processed experimental data in an open repository.
\end{itemize}
The remainder of this paper is organised as follows: Section~\ref{sec:methods} introduces the indices, and Section~\ref{sec:fe} presents the numerical case and the classification of its modes. Section~\ref{sec:exp} presents the two experimental cases, the XB-2 wing and the Sheraton Universal Hotel, and the classification of their modes. Section~\ref{sec:discussion} then discusses the results and their limitations, and the conclusions are drawn in Section~\ref{sec:conclusions}.

\section{Methods}\label{sec:methods}
This section introduces the four classification indices. The plate and the XB-2 wing, presented in Sections~\ref{sec:fe} and~\ref{sec:xb2}, share the same axes, whereas the building of Section~\ref{sec:sheraton} keeps its own axes, given in \cref{fig:sheraton}. On the plate and the wing, $x$ is chordwise, positive from the leading edge (LE) to the trailing edge (TE); $y$ is spanwise, positive outboard from the constrained end; and $z$ is positive upwards. On the building, $x$ points East, $y$ points North and $z$ points up, along its height.

Let a mode be sampled by $N$ sensors at positions $\mathbf{p}_i \in \mathbb{R}^3$, $i = 1, \dots, N$, each returning the motion vector $\mathbf{u}_i \in \mathbb{R}^3$ of that mode at the sensor. 
For experimental data, the sensor vectors are obtained from a real-valued representation
of each measured complex-valued mode (the adequacy of this assumption is assessed using the Modal Phase Collinearity \cite{Pappa1993} evaluated on the original complex modal vector before real-valued approximation, to discard physically meaningless identifications).
No direction is normalised separately, so that the relative amplitude of the components in different directions is preserved. {Since the four indices are invariant to a common scaling of the mode, the mode shapes should be scaled to a unit norm of their sensor vectors over the three axes, such that the numerical tolerances of an implementation act on a common scale.} The structure has a known span axis with unit direction $\hat{\mathbf{a}}$ passing through a reference point $\mathbf{p}_0$ at the constrained end.

Note that the classification rule is intended for global transverse bending and torsional modes of beam-like structures, including the lateral floor motions in high-rise shear-type buildings.
Axial, rigid-body, and local deformation modes are generally less relevant for the global analysis of this beam-like structures; thus, they fall outside of the present taxonomy, and must be excluded using structural or modal information independent of these indices.

The bending score $B$ measures how far the sensor vectors share one direction and one sign:
\begin{equation}
  B = \frac{\left\| \bar{\mathbf{u}} \right\|}{\frac{1}{N}\sum_{i=1}^{N} \left\| \mathbf{u}_i \right\|}, \qquad \bar{\mathbf{u}} = \frac{1}{N}\sum_{i=1}^{N} \mathbf{u}_i,
  \label{eq:B}
\end{equation}
where $\bar{\mathbf{u}}$ is the mean motion vector. $B = 1$ when all sensors move in phase along the same direction, as in a first bending mode, and $B$ decreases when parts of the structure move in opposition, across a nodal line or across the twist axis.

The torsion score $T$ describes the motion that remains once the common direction $\hat{\mathbf{d}} = \bar{\mathbf{u}} / \|\bar{\mathbf{u}}\|$ is removed. The residual of each sensor and the ideal tangential direction of a rotation about the span axis are
\begin{equation}
  \mathbf{r}_i = \mathbf{u}_i - \left( \mathbf{u}_i \cdot \hat{\mathbf{d}} \right) \hat{\mathbf{d}}, \qquad \mathbf{t}_i = \hat{\mathbf{a}} \times \left( \mathbf{p}_i - \mathbf{p}_0 \right).
  \label{eq:residual}
\end{equation}
and the score is the mean absolute cosine between them over the $N_r$ sensors with both a non-zero residual and a non-zero tangential vector:
\begin{equation}
  T = \frac{1}{N_r}
  \sum_{\substack{i:\,\|\mathbf{r}_i\|>0\\
                       \|\mathbf{t}_i\|>0}}
  \frac{|\mathbf{r}_i\cdot\mathbf{t}_i|}
       {\|\mathbf{r}_i\|\,\|\mathbf{t}_i\|}.
  \label{eq:T}
\end{equation}

$T$ approaches 1 when the residual motion is tangential to a rotation about the span axis, and is 0 when there is no residual left. 
Since the magnitude of $\mathbf{t}_i$ cancels in the cosine,
$T$ depends on the local tangential direction and the residual
displacement direction, without explicit weighting by radial
distance from the span axis.

The remaining two indices use sensors grouped in $N_s$ sections along the span axis, with one predefined transverse sensor pair per section.
The same pairs are retained for all modes of a given layout.
One valid pair is sufficient to compute $\tau$ for non-zero out-of-plane motion, whereas the section-wise comparability test requires at least two distinct paired sections.
The full classification rule is therefore applied only when at least two such sections are available.
The reference plane contains $\hat{\mathbf{a}}$ and the transverse direction of the sections, that is, the direction $\hat{\mathbf{e}}_s$ of the line that joins the pair of section $s$. Its unit normal $\hat{\mathbf{n}}$, nominally $\hat{\mathbf{a}} \times \hat{\mathbf{e}}_s / \| \hat{\mathbf{a}} \times \hat{\mathbf{e}}_s \|$, is prescribed as common to all sections: it is the vertical direction for the plate and the wing, and North for the building.
For non-coplanar pair directions, the sensor-separation vectors are projected onto this prescribed plane to define the transverse pair geometry; a non-zero transverse projection is required.
The motion along $\hat{\mathbf{n}}$ is called out-of-plane motion, and the motion within the reference plane in-plane motion. For section $s$, let $w^{(1)}_s$ and $w^{(2)}_s$ be the out-of-plane motion $\mathbf{u}_i \cdot \hat{\mathbf{n}}$ of its two sensors. 
The symmetric displacement component (heave),
$h_s = (w^{(2)}_s + w^{(1)}_s)/2$, and the antisymmetric
displacement component (twist),
{$\delta_s = (w^{(2)}_s - w^{(1)}_s)/2$},
define the twist share
\begin{equation}
  \tau = \frac{\sum_{s=1}^{N_s} \mbox{{$\delta_s$}}^2}{\sum_{s=1}^{N_s} \left( h_s^2 + \mbox{{$\delta_s$}}^2 \right)},
  \label{eq:tau}
\end{equation}

Both $h_s$ and {$\delta_s$} have displacement units{.}
For a rigid section undergoing a small rotation $\theta_s$ about the span axis, {$|\delta_s| = d_s|\theta_s|/2$}, where $d_s$ is the sensor separation projected onto the transverse direction within the reference plane.
Consequently, the squared  antisymmetric displacement introduces an implicit weighting of angular twist proportional to $d_s^2$ when contributions from different sections are combined.

The twist share is 0 for a mode in which the two sensors of every section move together, and 1 for a mode in which they move in opposition. Finally, the in-plane ratio compares the largest in-plane motion with the largest out-of-plane motion:
\begin{equation}
  \rho = \frac{\max_i \left\| \mathbf{u}_i - \left( \mathbf{u}_i \cdot \hat{\mathbf{n}} \right) \hat{\mathbf{n}} \right\|}{\max_i \left| \mathbf{u}_i \cdot \hat{\mathbf{n}} \right|},
  \label{eq:rho}
\end{equation}
where the numerator is the largest norm of the in-plane component. For a wing, $\hat{\mathbf{n}} = \hat{\mathbf{z}}$ and the numerator reduces to $\sqrt{u_{x,i}^2 + u_{y,i}^2}$. A mode with $\rho > 1$ moves more within the reference plane than out of it.

For non-degenerate inputs, $B$, $T$, and $\tau$ lie between 0 and 1,
whereas $\tau = 1/2$ and $\rho = 1$ are equality points. 
In particular, $\tau=0$ and $\tau=1$ correspond to entirely
symmetric and entirely antisymmetric sampled out-of-plane
displacements, respectively, with this last case suggesting torsion. 
An intermediate $\tau$ reflects a combination of the two motions. Within this range, $\tau = 1/2$ is the equal point of the twist and heave shares, both referring to the midpoint of each pair. Hence, $\tau$ is less than 1/2 when the heave share is larger, and greater than 1/2 when the twist share is larger. In the same way, $\rho$ is 0 for a purely out-of-plane mode and grows with the in-plane motion.

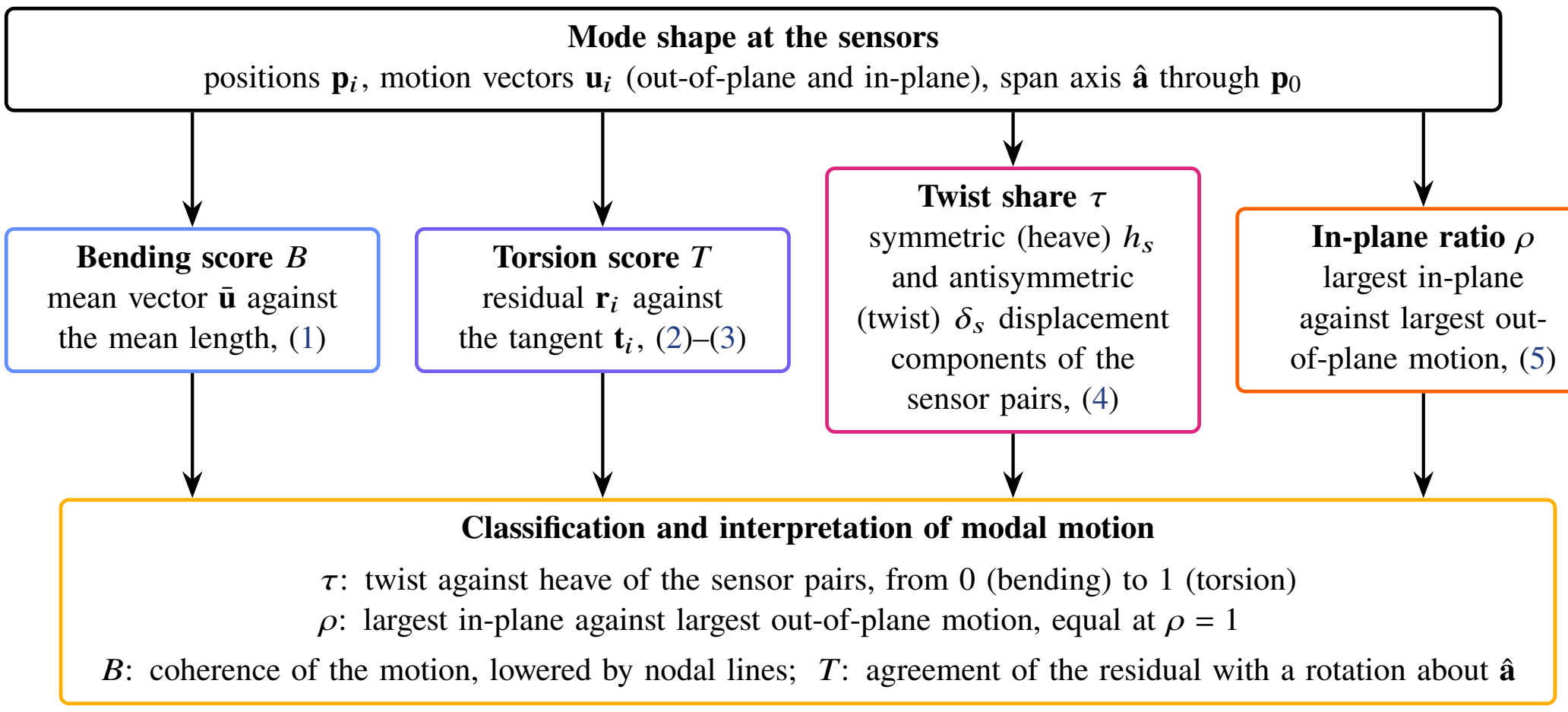
\begin{figure}[!t]
  \centering
  \definecolor{ibmblue}{HTML}{648FFF}\definecolor{ibmpurple}{HTML}{785EF0}
  \definecolor{ibmmagenta}{HTML}{DC267F}\definecolor{ibmorange}{HTML}{FE6100}\definecolor{ibmyellow}{HTML}{FFB000}
  \begin{tikzpicture}[>=Stealth, font=\small, node distance=10mm and 3mm,
      box/.style={draw, line width=0.9pt, rounded corners=2pt, align=center, inner sep=4pt}]
    \node[box, draw=black, text width=0.86\linewidth] (in) {\textbf{Mode shape at the sensors}\\ positions $\mathbf{p}_i$, motion vectors $\mathbf{u}_i$ (out-of-plane and in-plane), span axis $\hat{\mathbf{a}}$ through $\mathbf{p}_0$};
    \node[box, draw=ibmblue, text width=0.2\linewidth, below=of in.south west, anchor=north west] (B) {\textbf{Bending score} $B$\\ mean vector $\bar{\mathbf{u}}$ against the mean length, \eqref{eq:B}};
    \node[box, draw=ibmpurple, text width=0.2\linewidth, right=of B] (T) {\textbf{Torsion score} $T$\\ residual $\mathbf{r}_i$ against the tangent $\mathbf{t}_i$, \eqref{eq:residual}--\eqref{eq:T}};
    \node[box, draw=ibmmagenta, text width=0.2\linewidth, right=of T] (tau) {\textbf{Twist share} $\tau$\\ symmetric (heave) $h_s$ and antisymmetric (twist) {$\delta_s$} displacement components of the sensor pairs, \eqref{eq:tau}};
    \node[box, draw=ibmorange, text width=0.2\linewidth, right=of tau] (rho) {\textbf{In-plane ratio} $\rho$\\ largest in-plane against largest out-of-plane motion, \eqref{eq:rho}};
    \node[box, draw=ibmyellow, text width=0.86\linewidth, anchor=north] (out) at ($($(B.south west)!0.5!(rho.south east)$) + (0,-10mm)$) {\textbf{Classification and interpretation of modal motion}\\[3pt]
      \begin{tabular}{@{}l@{}}
        $\tau$: twist against heave of the sensor pairs, from 0 (bending) to 1 (torsion) \\
        $\rho$: largest in-plane against largest out-of-plane motion, equal at $\rho = 1$
      \end{tabular}\\[3pt]
      $B$: coherence of the motion, lowered by nodal lines; \ $T$: agreement of the residual with a rotation about $\hat{\mathbf{a}}$};
    \foreach \n in {B, T, tau, rho} {\draw[->, line width=0.8pt] (in.south -| \n.north) -- (\n.north); \draw[->, line width=0.8pt] (\n.south) -- (\n.south |- out.north);}
  \end{tikzpicture}
  \caption{\captiontitle{Classification process.} The mode shape sampled at the sensors gives four indices, which are read together to \protect{characterise} the modal motion; the rule that assigns the modal label from $\tau$ and $\rho$ is defined in Section~\ref{sec:discussion}.}
  \label{fig:flowchart}
\end{figure}

On the other hand, a low value of $B$ does not indicate torsion by itself, because nodal sign reversals in higher bending modes can also reduce the mean motion vector. When $\|\bar{\mathbf{u}}\|$ approaches zero, the common direction $\hat{\mathbf{d}}$ becomes poorly defined and $T$ is ill-conditioned. The classification process is summarised in \cref{fig:flowchart}.

To summarise, in the context of this paper, ``coupled'' denotes the coexistence of motion components in the sampled mode shape according to the stated classification rule. The terms ``dominant'' and ``secondary'' refer to ordering under the {displacement}-based criteria for the specified sensor layout, available components, and reference axes. For the comparison between in-plane and out-of-plane motion, ``dominance'' is defined by the maximum sampled displacement amplitudes, which may occur at different sensors. Therefore, dominance is intended in terms of peak displacement for the specified array, without implying energy dominance.

\section{Numerical case: tapered plate with a swept trailing edge}\label{sec:fe}
\begin{figure}[!htp]
  \centering
  \includegraphics[width=0.9\linewidth]{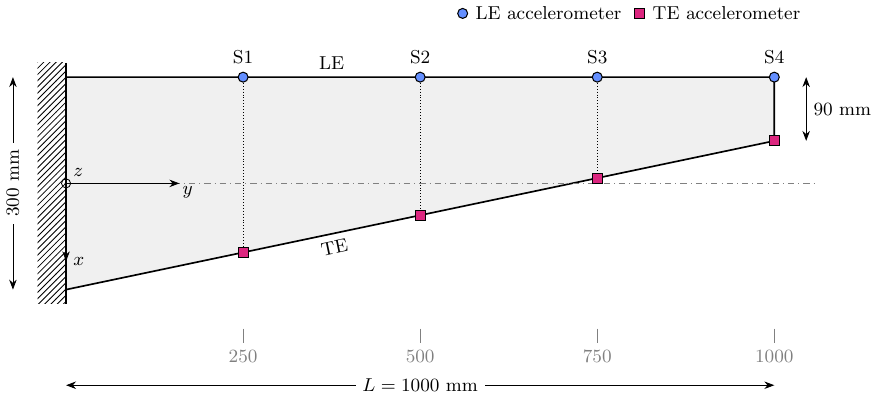}
  \caption{\captiontitle{Planform of the tapered plate and accelerometer stations.} The root is clamped; the origin of the axes is at the root mid-chord, with $z$ out of the page. Circles: LE accelerometers; squares: TE accelerometers.}
  \label{fig:plate}
\end{figure}

The numerical case is a cantilever plate of span $L =$ 1000 mm, root chord 300 mm, tip chord 90 mm and thickness $t =$ 2 mm, shown in \cref{fig:plate}. Its LE is straight and its TE is swept. Hence, the plate is not symmetric about the span axis, and its modes couple bending and torsion, resembling a low-speed cantilever wing. The material is 6082-T6 aluminium, with a Young's Modulus of 70 GPa, a Poisson ratio of 0.33 and a density of 2700 kg m\textsuperscript{-3}.

The plate is modelled in Ansys Mechanical APDL 2026 R1 with four-node SHELL181 elements in a mapped mesh. All degrees of freedom of the root nodes are constrained. The mesh has 800 divisions along the span and 156 along the chord, with a mean element size of 1.25 mm, for a total of 125,757 nodes\footnote{This was the finest mesh allowed by the licence used.}. The modes are extracted with the Block Lanczos solver.

Eight sensors are placed at the LE and at the TE of four stations, $y =$ 250, 500, 750 and 1000 mm. Since a shell element returns the motion of its mid-surface, the motion of each sensor is evaluated at the top surface of the plate as follows:
\begin{equation}
  \mathbf{u}^{\mathrm{top}} = \mathbf{u}^{\mathrm{mid}} + \boldsymbol{\theta} \times \left( \frac{t}{2}\, \hat{\mathbf{z}} \right),
  \label{eq:top}
\end{equation}
where $\mathbf{u}^{\mathrm{mid}}$ and $\boldsymbol{\theta}$ are the nodal displacement and rotation vectors of the mode and $\hat{\mathbf{z}}$ is the normal to the plate. Both fields are interpolated linearly onto a 1 mm grid, and each sensor is assigned to the nearest grid point inside the planform, at most 0.5 mm from its nominal position. 
After sampling on the 1~mm grid, the projected LE--TE sensor
separations at the four stations are 247, 195, 142 and 90~mm.
The span axis is $\hat{\mathbf{a}} = \hat{\mathbf{y}}$, through the root mid-chord. 
Since the TE is swept, the TE sensors of stations 3 and 4 lie on the LE side of this axis. This does not affect $\tau$, which is referred to the midpoint of each pair, whereas $T$ is referred to the span axis. The first eight modes are classified.

\begin{figure}[!htp]
  \centering
  \begin{subfigure}[b]{0.4\linewidth}\centering\includegraphics[width=\linewidth]{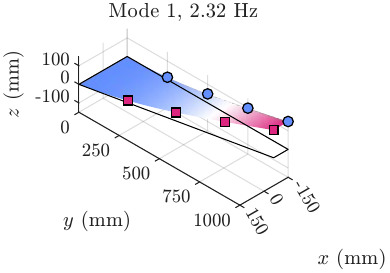}\caption{}\end{subfigure}
  \begin{subfigure}[b]{0.4\linewidth}\centering\includegraphics[width=\linewidth]{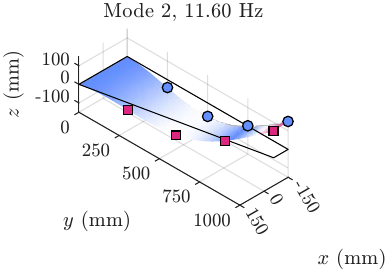}\caption{}\end{subfigure}

  \begin{subfigure}[b]{0.4\linewidth}\centering\includegraphics[width=\linewidth]{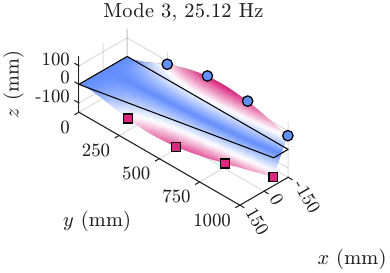}\caption{}\end{subfigure}
  \begin{subfigure}[b]{0.4\linewidth}\centering\includegraphics[width=\linewidth]{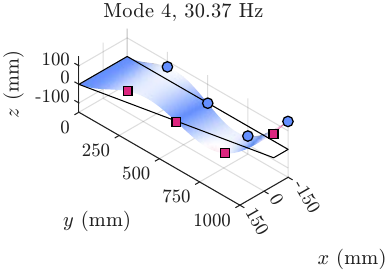}\caption{}\end{subfigure}

  \begin{subfigure}[b]{0.4\linewidth}\centering\includegraphics[width=\linewidth]{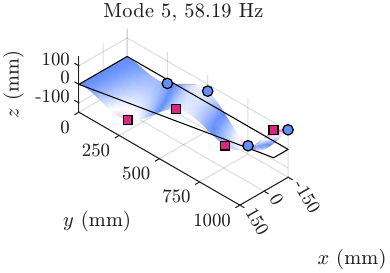}\caption{}\end{subfigure}
  \begin{subfigure}[b]{0.4\linewidth}\centering\includegraphics[width=\linewidth]{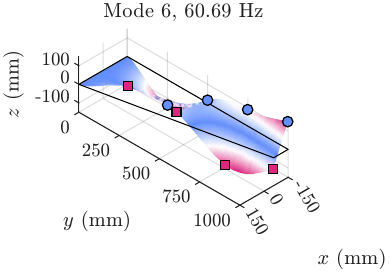}\caption{}\end{subfigure}

  \begin{subfigure}[b]{0.4\linewidth}\centering\includegraphics[width=\linewidth]{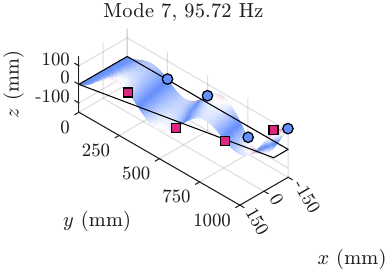}\caption{}\end{subfigure}
  \begin{subfigure}[b]{0.4\linewidth}\centering\includegraphics[width=\linewidth]{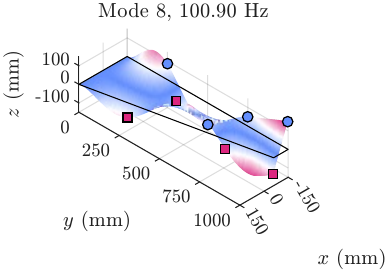}\caption{}\end{subfigure}

  \begin{subfigure}[b]{0.8\linewidth}\centering\includegraphics[width=\linewidth]{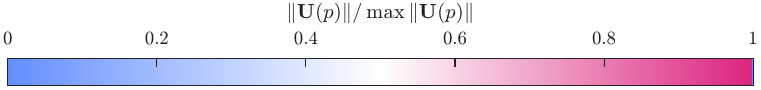}\end{subfigure}
  \caption{\captiontitle{First eight modes of the tapered plate with a swept trailing edge.} Isometric view of the top surface, displaced by the mode shape. Circles: LE accelerometers; squares: TE accelerometers. Note: For visualisation, the normalised modal displacements are multiplied by a display scale of 150~mm; the resulting amplitudes are arbitrary and do not represent physical vibration amplitudes.}
  \label{fig:femodes}
\end{figure}

The first eight modes of the plate lie between 2.32 Hz and 100.90 Hz. These modes are shown in \cref{fig:femodes}, while their indices are reported in \cref{tab:fe}. The type of each mode is assigned by inspecting the mode shapes in \cref{fig:femodes}, and the indices are compared with this assignment. All eight modes are out-of-plane. Consequently, the in-plane ratio is below 0.02 for all modes.

\begin{table}[!htp]
  \footnotesize
  \caption{\captiontitle{Classification indices of the first eight modes of the plate.} $B$: bending score; $T$: torsion score; $\tau$: twist share; $\rho$: in-plane ratio.}
  \label{tab:fe}
  \begin{oretablec}{|>{\bfseries}l|l|l|l|l|l|l|}
    \tablehead \textbf{Mode} & \textbf{$f_n$ (Hz)} & \textbf{$B$} & \textbf{$T$} & \textbf{$\tau$} & \textbf{$\rho$} & \textbf{Type} \\ \hline
    1 & 2.32 & 1.000 & 0.007 & 0.000 & 0.001 & out-of-plane bending \\ \hline
    2 & 11.60 & 0.103 & 0.030 & 0.002 & 0.004 & out-of-plane bending \\ \hline
    3 & 25.12 & 0.021 & 0.659 & 0.996 & 0.014 & torsion \\ \hline
    4 & 30.37 & 0.507 & 0.015 & 0.009 & 0.007 & out-of-plane bending \\ \hline
    5 & 58.19 & 0.202 & 0.037 & 0.101 & 0.010 & out-of-plane bending coupled with torsion \\ \hline
    6 & 60.69 & 0.079 & 0.086 & 0.900 & 0.016 & torsion coupled with out-of-plane bending \\ \hline
    7 & 95.72 & 0.539 & 0.011 & 0.141 & 0.013 & out-of-plane bending coupled with torsion \\ \hline
    8 & 100.90 & 0.084 & 0.083 & 0.971 & 0.017 & torsion \\ \hline
  \end{oretablec}
\end{table}

The twist share agrees with these types. Modes 1, 2 and 4 are bending modes, with $\tau$ not above 0.009, and modes 3 and 8 are torsion modes, with $\tau$ of 0.996 and 0.971. Modes 5 and 6 form a pair 2.5 Hz apart. 
Both modes contain bending- and torsion-like displacement components.
Specifically, mode 5 is a bending mode with a twist share of 0.101, and mode 6 is a torsion mode with a twist share of 0.900. 
The fifth predominantly bending mode (mode 7) has a twist share of 0.141, whereas the third predominantly torsional mode (mode 8) has a twist share of 0.971.
The bending score separates the first bending mode, $B = 1$, from the higher modes, whose nodal lines reverse the motion of part of the plate and lower $B$ to between 0.021 and 0.539. On the other hand, the torsion score is not above 0.037 for the bending modes and reaches 0.659 for the first torsion mode. However, for the second and third torsion modes, the torsion score is below 0.09.

\section{Experimental case studies}\label{sec:exp}
\subsection{The eXperimental BeaRDS 2 flexible wing}\label{sec:xb2}
The XB-2 is a flexible wing model of a conceptual high-aspect-ratio wing for an aircraft of the Airbus A320 class,  dynamically scaled at 1:16. The XB-2 was designed within the BeaRDS programme at Cranfield University\footnote{At the time of writing, Cranfield University is expected to be soon renamed as Cranfield Institute of Technology at King's College London.}.\cite{Yusuf2019} It has a nominal overall semi-span of 1.5~m, a National Advisory Committee for Aeronautics (NACA) 23015 aerofoil,\cite{Abbott1945} a mean aerodynamic chord of 172 mm, a taper ratio of 0.35 and no sweep of the quarter-chord line.\cite{Dessena2022b} 
The free length of the main spar outboard of the clamp is 1.325~m, which defines the structural spanwise coordinate used here.\cite{Dessena2022b}
Its structure comprises a 6082-T6 aluminium spar, a stainless steel stiffening tube and a skin printed in Digital ABS (acrylonitrile butadiene styrene) and Agilus 30. The full wing is clamped on an electrodynamic shaker. The excitation is a random base acceleration limited to 2--400 Hz, for 20 min. The record used here is acquired on the same specimen as the original benchmark,\cite{Dessena2022b} whose dataset is openly available,\footnote{G. Dessena, Data supporting: 'Ground Vibration Testing of a Flexible Wing: A Benchmark and Case Study'. Cranfield Online Research Data (CORD), 2022. \url{https://doi.org/10.17862/cranfield.rd.19077023}} but at a different amplitude.

\begin{figure}[!htp]
  \centering
  \includegraphics[width=\linewidth]{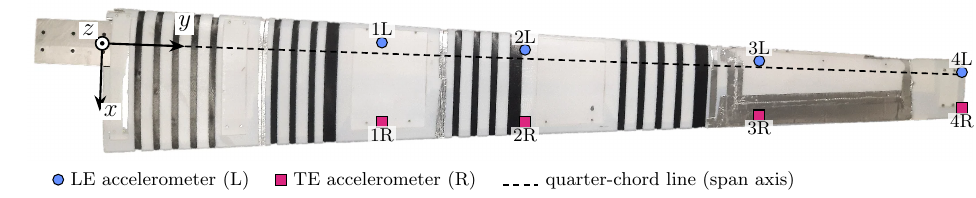}
  \caption{\captiontitle{XB-2 wing with the accelerometer positions and the axes.} The photograph is reproduced, mirrored, from Dessena et al.\cite{Dessena2022b} under a Creative Commons Attribution (CC BY) 4.0 licence; the root is on the left and the LE at the top. The dashed line is the quarter-chord line, taken as the span axis, and the origin of the axes is at the clamp exit, at the fixed end of the exposed spar, with $z$ out of the page.}
  \label{fig:xb2sensors}
\end{figure}

The response is measured by eight accelerometers at the LE and the TE of four stations, while a further accelerometer on the clamp measures the input. 
The stations are at $y=430$, 650, 1010, and 1325~mm, measured from the clamp exit at the fixed end of the exposed spar. The last station corresponds to the free spar tip in this structural coordinate system.
At each station, the accelerometers are placed as close as possible to the LE and to the TE, as shown in \cref{fig:xb2sensors}. The distances from the LE, from 27.5 mm at the first station to 18 mm at the tip, were measured during the test. The distances from the TE range from 28 mm to 14 mm.

Each measuring point is placed on the upper surface of the NACA 23015 profile at its chord fraction, raised by half the height of the accelerometer, since the sensing element lies at mid-height of its case. The heights are taken from the manufacturer datasheets of the accelerometer models used in the test, and the resulting positions are reported in \cref{tab:xb2sensors}.

\begin{table}[!htp]
  \footnotesize
  \caption{\captiontitle{Position of the XB-2 accelerometers.} $x$ from the quarter-chord line, positive towards the TE; $y$ from the clamp root; $z$ is the height of the measuring point, given by the upper surface of the aerofoil plus half the height of the accelerometer. Ch.: measured directions.}
  \label{tab:xb2sensors}
  \begin{oretablec}{|>{\bfseries}l|l|l|l|l|l|}
    \tablehead \textbf{Sensor} & \textbf{Model} & \textbf{$x$ (mm)} & \textbf{$y$ (mm)} & \textbf{$z$ (mm)} & \textbf{Ch.} \\ \hline
    1R (TE) & PCB 356A16 & 105 & 430 & 12 & $x$, $z$ \\ \hline
    1L (LE) & Isotron 7251A & $-$17 & 430 & 20 & $z$ \\ \hline
    2R (TE) & PCB 356A16 & 97 & 650 & 11 & $x$, $z$ \\ \hline
    2L (LE) & Isotron 7251A & $-$14 & 650 & 18 & $z$ \\ \hline
    3R (TE) & PCB 356A45 & 74 & 1010 & 8 & $x$, $z$ \\ \hline
    3L (LE) & Isotron 7251A & $-$10 & 1010 & 15 & $z$ \\ \hline
    4R (TE) & B\&K 4507-002 & 51 & 1325 & 8 & $z$ \\ \hline
    4L (LE) & B\&K 4507-002 & $-$4 & 1325 & 13 & $z$ \\ \hline
  \end{oretablec}
\end{table}

The projected LE--TE sensor separations used in the calculation
are 121.8, 111.1, 83.7 and 55.2~mm at stations 1 to 4.
The vertical channel of every accelerometer gives the $z$ component of $\mathbf{u}_i$, and the chordwise channel of the triaxial accelerometers at stations 1 to 3 gives the $x$ component at the TE. The components that are not measured are set to zero. This choice leaves the twist share unaffected, since $\tau$ uses only the vertical motion, which every accelerometer measures. Nor can it raise the in-plane ratio, since $\rho$ compares maxima; thus, $\rho$ is a lower bound, which an unmeasured chordwise component could only increase. On the other hand, $B$ and $T$ depend on this choice, because the accelerometers do not all measure the same components.
The acquisition-axis labels are mapped as $Y\mapsto z$ and $Z\mapsto x$.
The released channels, expressed in $g$, are used without additional sign changes or channel-dependent rescaling
in the present analysis.
Since the quarter-chord line of the XB-2 is not swept, it is taken as the span axis, $\hat{\mathbf{a}} = \hat{\mathbf{y}}$, through the clamp root.

The modes are identified as follows. First, the signals, sampled at 5120 Hz, are decimated to 256 Hz and detrended. 
Dimensionless acceleration transmissibilities from the clamp reference to eleven wing channels are estimated with the $H_1$ estimator, using Hann windows of 2048 samples with 50\% overlap.
The modal parameters are extracted in the 2--30 Hz band with the real-valued iLF,\footnote{G. Dessena, A tutorial for the improved Loewner Framework for modal analysis. Department of Aerospace Engineering, Universidad Carlos III de Madrid, 16 Nov., 2024. doi:\url{https://doi.org/10.5281/zenodo.13863292}.} for even model orders from 2 to 80.  
Stable poles are retained for damping ratios between 0.001 and 0.3, relative variations below 1\% in frequency and 10\% in damping between consecutive model orders, and MAC value greater than 0.95 with respect to the paired (i.e., most similar) modal vector from the preceding order.
The physical modes are then selected by inspection of the stabilisation diagram.

\begin{figure}[!t]
  \centering
  \begin{subfigure}[b]{0.4\linewidth}\centering\includegraphics[width=\linewidth]{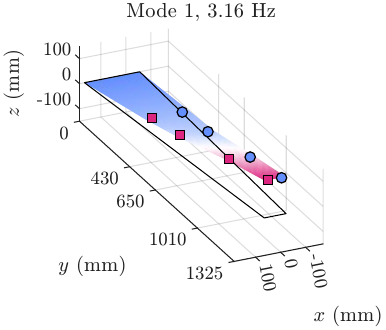}\caption{}\end{subfigure}
  \begin{subfigure}[b]{0.4\linewidth}\centering\includegraphics[width=\linewidth]{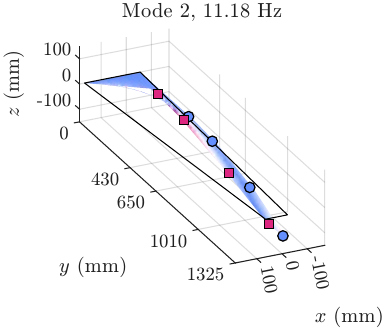}\caption{}\end{subfigure}

  \begin{subfigure}[b]{0.4\linewidth}\centering\includegraphics[width=\linewidth]{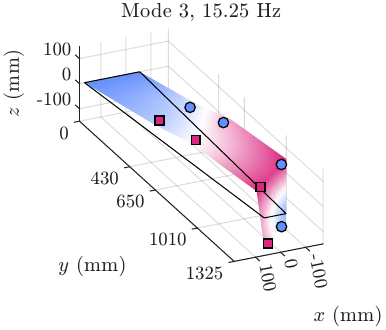}\caption{}\end{subfigure}
  \begin{subfigure}[b]{0.4\linewidth}\centering\includegraphics[width=\linewidth]{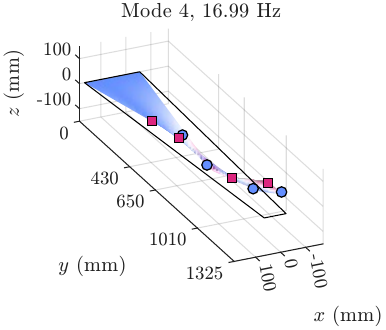}\caption{}\end{subfigure}

  \begin{subfigure}[b]{0.8\linewidth}\centering\includegraphics[width=\linewidth]{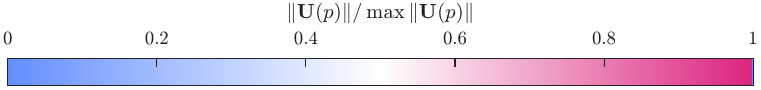}\end{subfigure}
  \caption{\captiontitle{Identified modes of the XB-2 wing.} Isometric view of the planform displaced by the mode shape, interpolated between the accelerometers.
  Circles: LE accelerometers; squares: TE accelerometers. Note: For visualisation, the normalised modal displacements are multiplied by a display scale of 150~mm; the resulting amplitudes are arbitrary and do not represent physical vibration amplitudes.}
  \label{fig:xb2modes}
\end{figure}

Four modes are selected, at 3.16, 11.18, 15.25 and 16.99 Hz. These modes are shown in \cref{fig:xb2modes}. Three of them correspond to the first bending mode and to the two coupled modes reported for the high-input case of the full wing,\cite{Dessena2022b} at 3.139, 11.196 and 16.988 Hz. Their frequencies agree with these within 0.7\%, although the control level of the record, 0.35 g root mean square (RMS), is lower than the 1.712 g RMS of that case. The remaining mode, at 15.25 Hz, has no counterpart among the modes reported for the XB-2, since the original identification relied on the vertical channels only.

\begin{table}[!htp]
  \footnotesize
  \caption{\captiontitle{Classification indices of the identified modes of the XB-2 wing.} $\zeta_n$: damping ratio; $B$: bending score; $T$: torsion score; $\tau$: twist share; $\rho$: in-plane ratio. For modes 2 and 4, the order of the dominant and the secondary motion follows the equality point $\tau = 1/2$, whose limits on this layout are discussed in Section~\ref{sec:discussion}.}
  \label{tab:xb2}
  \begin{oretablec}{|>{\bfseries}l|l|l|l|l|l|l|>{\raggedright\arraybackslash}p{4.2cm}|}
    \tablehead \textbf{Mode} & \textbf{$f_n$ (Hz)} & \textbf{$\zeta_n$} & \textbf{$B$} & \textbf{$T$} & \textbf{$\tau$} & \textbf{$\rho$} & \textbf{Type} \\ \hline
    1 & 3.16 & 0.031 & 0.996 & 0.466 & 0.001 & 0.10 & out-of-plane bending \\ \hline
    2 & 11.18 & 0.057 & 0.417 & 0.608 & 0.451 & 0.25 & out-of-plane bending coupled with torsion \\ \hline
    3 & 15.25 & 0.030 & 0.453 & 0.729 & 0.264 & 1.33 & in-plane bending coupled with out-of-plane bending and torsion \\ \hline
    4 & 16.99 & 0.044 & 0.235 & 0.604 & 0.339 & 0.20 & out-of-plane bending coupled with torsion \\ \hline
  \end{oretablec}
\end{table}

As reported in \cref{tab:xb2}, the first mode is a bending mode, with $B = 0.996$ and $\tau = 0.001$. The second and fourth modes are coupled, in agreement with the labels of the original benchmark.\cite{Dessena2022b} Their twist shares are 0.451 and 0.339. The order of their dominant and secondary motion follows $\tau = 1/2$, whose limits on this layout are discussed in Section~\ref{sec:discussion}. Hence, the coupling of these two modes is supported, since both twist shares lie well above $\tau_b$. The order of their two components, instead, depends on the reference for $\tau$; with half the pure-rotation value as the threshold, torsion dominates in the second mode about both reference lines, and in the fourth about the quarter-chord line only. 
Following modal identification, the  third mode at 15.25~Hz is classified
as in-plane dominant, according to this peak-amplitude criterion, since $\rho=1.33$ indicates that the maximum measured chordwise amplitude exceeds the maximum measured vertical amplitude.
The twist share of its vertical component is 0.264. At stations 1 to 3, where the chordwise motion is measured, the ratio of the largest chordwise to the largest vertical motion is 0.98, 1.10 and 3.70, against 0.13, 0.08 and 0.75 in the fourth mode. To the authors' knowledge, this is the first identification of the modal parameters of an in-plane mode of the XB-2 wing. The earlier campaigns of this wing note a lag-dominated mode without investigating it.\cite{Dessena2022h} The torsion score lies between 0.466 and 0.729 for all four modes, since the chordwise channels at the TE give a residual for every mode, including the first bending mode. Hence, on the XB-2, the type of each mode is successfully classified by the proposed indices, based on the measured components.

\subsection{Sheraton Universal Hotel}\label{sec:sheraton}
\begin{figure}[!htp]
  \centering
  \begin{subfigure}[b]{0.58\linewidth}\centering\includegraphics[width=\linewidth]{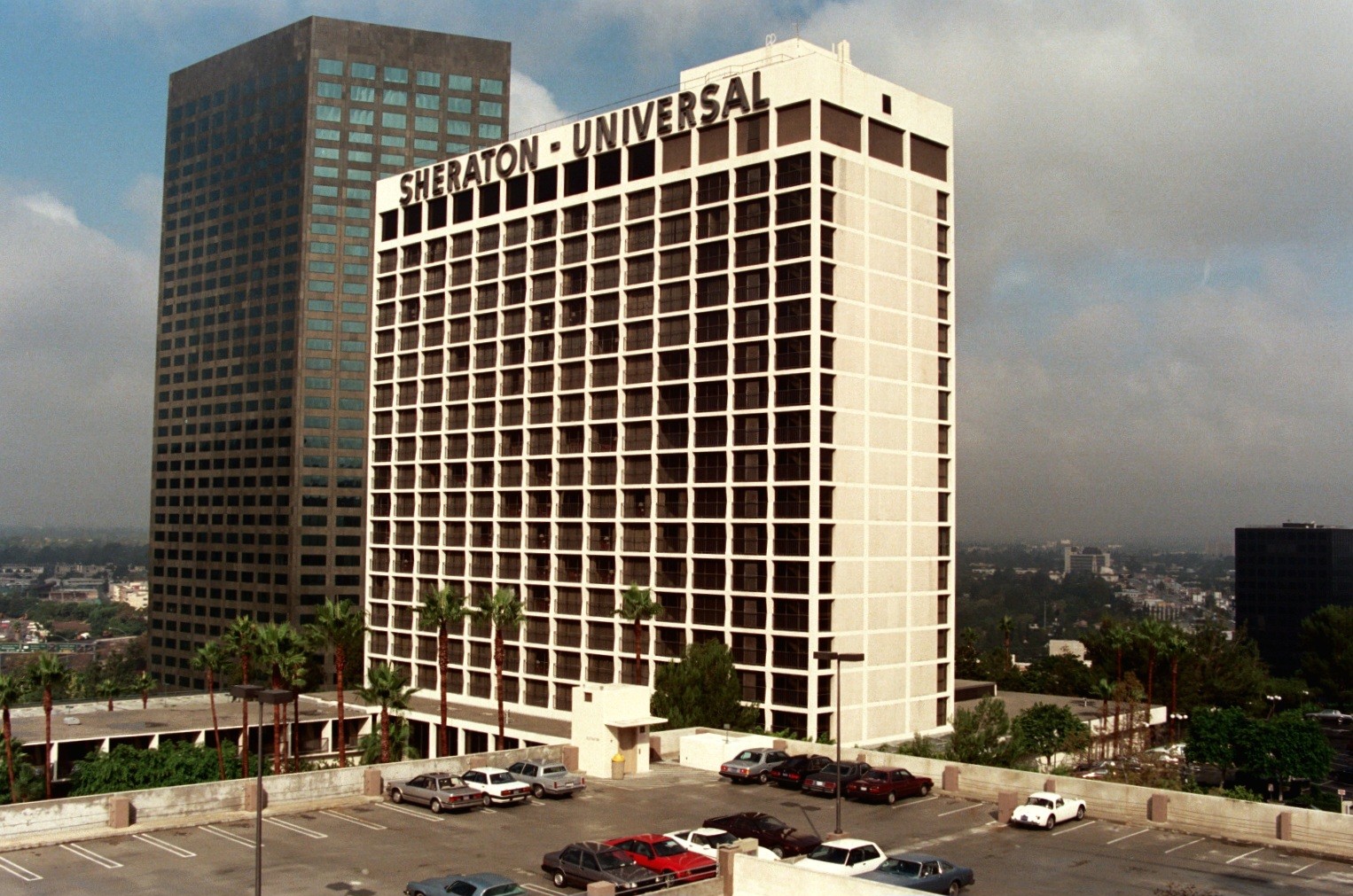}\caption{}\end{subfigure}\hspace{2mm}
  \begin{subfigure}[b]{0.37\linewidth}\centering\includegraphics[width=\linewidth]{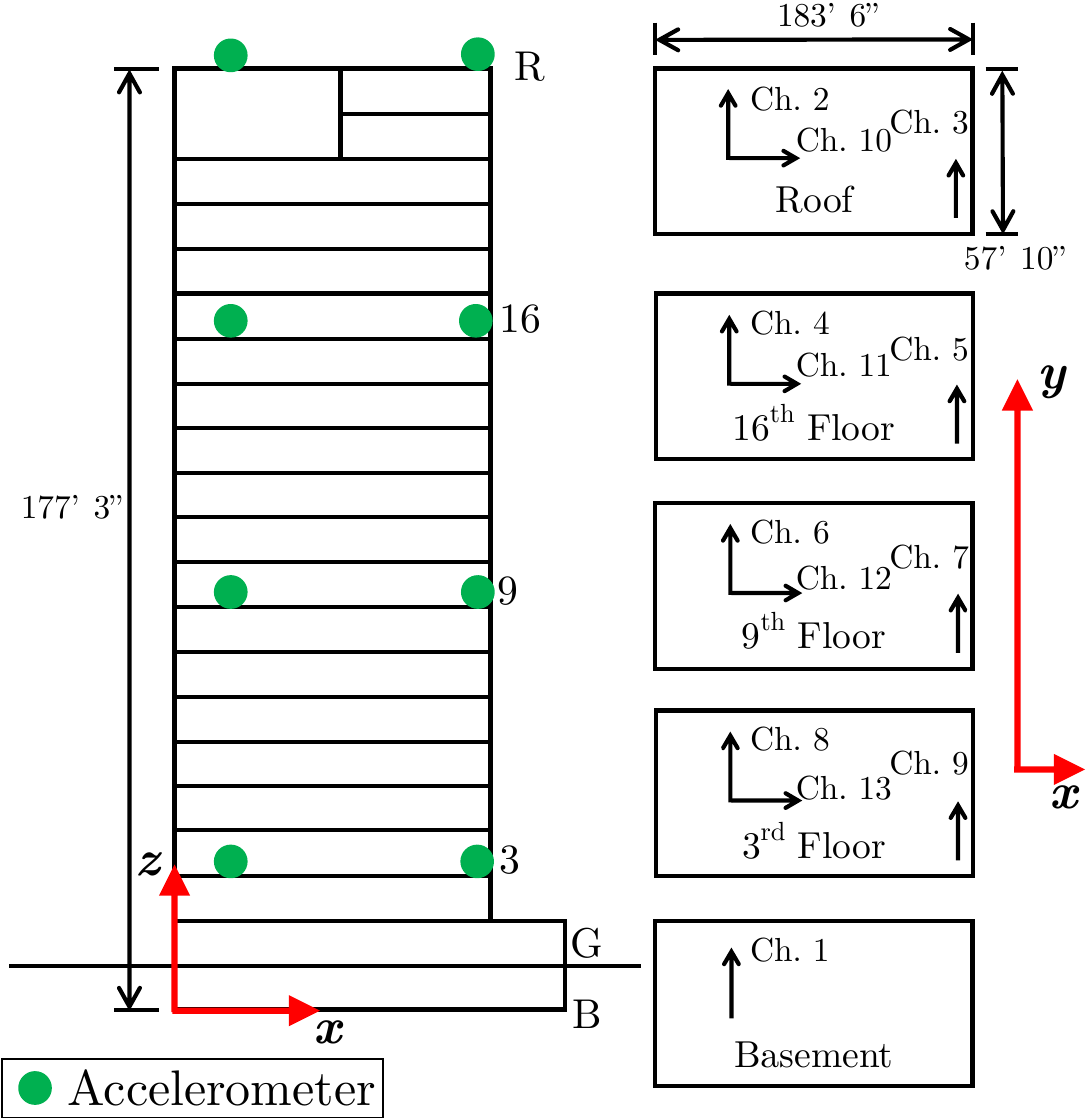}\caption{}\end{subfigure}
  \caption{\captiontitle{Sheraton Universal Hotel and its accelerometers.} (a) The building in 1987; (b) elevation and floor plans with the accelerometer channels and the axes, $x$ East, $y$ North and $z$ up. Both are reproduced from Dessena et al.\cite{Dessena2024f} under a CC BY licence; the photograph is adapted there from Flickr under a CC BY licence.}
  \label{fig:sheraton}
\end{figure}

\begin{figure}[!htp]
  \centering
  \begin{subfigure}[b]{0.32\linewidth}\centering\includegraphics[width=\linewidth]{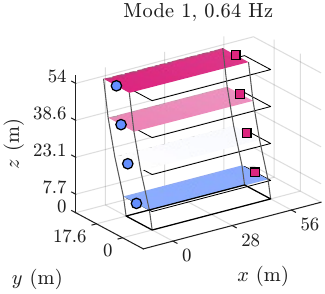}\caption{}\end{subfigure}
  \begin{subfigure}[b]{0.32\linewidth}\centering\includegraphics[width=\linewidth]{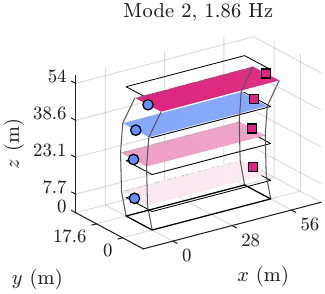}\caption{}\end{subfigure}
  \begin{subfigure}[b]{0.32\linewidth}\centering\includegraphics[width=\linewidth]{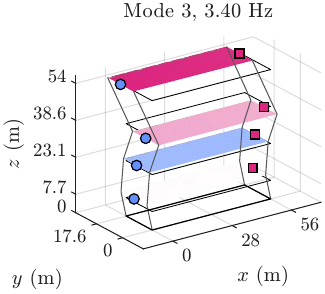}\caption{}\end{subfigure}\\[2mm]
  \begin{subfigure}[b]{0.32\linewidth}\centering\includegraphics[width=\linewidth]{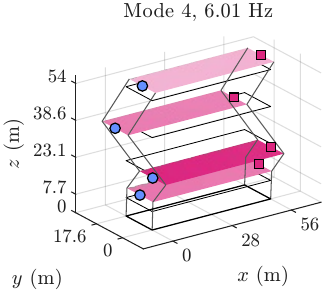}\caption{}\end{subfigure}
  \begin{subfigure}[b]{0.32\linewidth}\centering\includegraphics[width=\linewidth]{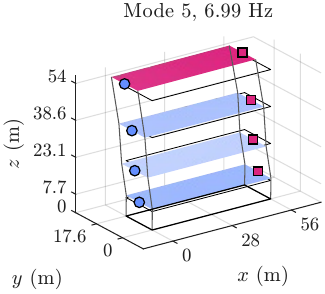}\caption{}\end{subfigure}\\[2mm]
  \begin{subfigure}[b]{0.8\linewidth}\centering\includegraphics[width=\linewidth]{Figure5e.pdf}\end{subfigure}
  \caption{\captiontitle{Mode shapes of the Sheraton Universal Hotel.} Isometric view of the instrumented floors, each displaced as a rigid plate by the translation and the rotation read from the published mode shapes.\cite{Dessena2024f} Circles: West North--South sensor points; squares: East North--South sensor points. Note: For visualisation, the normalised modal displacements are multiplied by a display scale of 12~m; the resulting amplitudes are arbitrary and do not represent physical vibration amplitudes.}
  \label{fig:sheratonmodes}
\end{figure}

The Sheraton Universal Hotel, in Universal City, North Hollywood, California (USA), is a reinforced concrete high-rise building, designed in 1967 and constructed the following year. Its general geometry is defined by a height of 54 m (177'3") with typical upper-floor dimensions of 17.6 m (57'10") by 56 m (183'6"). Ductile moment-resisting frames provide lateral resistance.\cite{Dessena2024f}
Its ambient vibration response, recorded on 18 December 1997, is measured on the 3rd, 9th and 16th floors and on the roof, each with two North--South accelerometers and one East--West accelerometer, as shown in \cref{fig:sheraton}.\cite{Dessena2024f} Five modes are identified from these records through NExT-LF, which pairs the Natural Excitation Technique (NExT) with the Loewner Framework (LF),\cite{Dessena2024f} and their mode shapes are openly available.\footnote{G. Dessena, Data supporting: NExT-LF: A Novel Operational Modal Analysis Method via Tangential Interpolation. Universidad Carlos III de Madrid, May 13, 2025. doi: \url{https://doi.org/10.5281/zenodo.14975925}.} The mode shape figures of this dataset give the translation and the rotation of each instrumented floor, taken as rigid in its plane. Each mode is sampled from these figures at the two North--South sensor points of every floor, taken at the ends of the 56 m side on the centreline of the plan. The five mode shapes sampled in this way are shown in \cref{fig:sheratonmodes}.

This case is an application of the indices to a civil engineering structure. The indices of Section~\ref{sec:methods} apply to the building with the height as the span axis, through the centroid of the plan at the base. The two North--South sensor points of each floor form the pair of each section. Hence, the reference plane is the vertical plane along the 56 m side and its normal $\hat{\mathbf{n}}$ points North, so that $\tau$ measures the twist of each floor about the vertical axis. Since the vertical motion is measured at no point, setting it to zero is equivalent to omitting it, and $\rho$ compares the largest East--West with the largest North--South motion.

\begin{table}[!htp]
  \footnotesize
  \caption{\captiontitle{Classification indices of the modes of the Sheraton Universal Hotel.} $f_n$ and $\zeta_n$ as identified through NExT-LF;\cite{Dessena2024f} $B$: bending score; $T$: torsion score; $\tau$: twist share; $\rho$: in-plane ratio, East--West against North--South motion.}
  \label{tab:sheraton}
  \begin{oretablec}{|>{\bfseries}l|l|l|l|l|l|l|>{\raggedright\arraybackslash}p{4.2cm}|}
    \tablehead \textbf{Mode} & \textbf{$f_n$ (Hz)} & \textbf{$\zeta_n$} & \textbf{$B$} & \textbf{$T$} & \textbf{$\tau$} & \textbf{$\rho$} & \textbf{Type} \\ \hline
    1 & 0.64 & 0.03 & 1.000 & 0.219 & 0.000 & 0.22 & North--South bending \\ \hline
    2 & 1.86 & 0.01 & 0.150 & 0.163 & 0.000 & 0.33 & North--South bending \\ \hline
    3 & 3.40 & 0.02 & 0.423 & 0.064 & 0.000 & 0.21 & North--South bending \\ \hline
    4 & 6.01 & 0.02 & 0.248 & 0.347 & 0.000 & 0.53 & North--South bending coupled with East--West bending \\ \hline
    5 & 6.99 & 0.01 & 0.677 & 0.215 & 0.000 & 0.27 & North--South bending coupled with East--West bending \\ \hline
  \end{oretablec}
\end{table}

As reported in \cref{tab:sheraton}, $\tau$ stays below 0.0005 in all five modes. 
The sampled ends therefore have nearly equal North–South modal displacements, moving basically together.
All five modes are thus bending modes in the North--South direction. The in-plane ratio ranges from 0.21 to 0.53, that is, the largest East--West motion is 21\% to 53\% of the largest North--South motion. 
In the first mode, the ratio of the East--West to the North--South motion lies between 0.11 and 0.24 at the four floors, consistently with the skew of its direction from North--South noted in the original identification.\cite{Dessena2024f} On the other hand, the East--West motion of the fourth mode is comparable to its North--South motion at the 3rd and 16th floors and at the roof; therefore, this mode is classified as North--South bending coupled with East--West bending. The ratio of the East--West to the North--South motion is 0.75, 0.16, 0.85 and 0.96 at the 3rd, 9th and 16th floors and at the roof, against 0.35, 0.37, 0.66 and 0.33 in the second mode. 
In the fifth mode, the largest North--South motion, on the roof, is about four times the largest East--West motion ($\rho =$ 0.27), whereas the East--West motion exceeds the North--South motion at the 9th and 16th floors, with ratios of 1.95 and 1.41 against 0.74 and 0.10 at the 3rd floor and on the roof. Accordingly, this mode is classified as North--South bending coupled with East--West bending.
As in the plate and on the XB-2, the first bending mode has the highest bending score, $B =$ 1.000. 
The torsion score lies between 0.064 and 0.347; since the two sensor points of each floor lie on a line through the span axis, it compares the residual with the North--South direction only.

\section{Discussion}\label{sec:discussion}
In the three case studies, $\tau$ quantifies the relative contribution of antisymmetric out-of-plane displacement within the sensor pairs, while $\rho$ compares the largest in-plane and out-of-plane amplitudes. On the plate, the resulting classifications agree with those assigned by visual inspection of the mode shapes.
In all three cases, they are complemented by a section-wise check of the in-plane motion, defined below. 
Across the three cases, $B$ takes its highest value for the first bending mode, while the behaviour of $T$ differs markedly between the plate and the XB-2 wing.
On the plate, measured on its top surface, the torsion score marks the first torsion mode but not the higher torsion modes. In these modes, the mean motion vector is close to vertical, so that the residual of each sensor is nearly chordwise. Since the tangential direction of a rotation about the span axis is nearly vertical at the sensors, the torsion score is low. On the XB-2, the torsion score is high for all four modes because the chordwise channels at the TE leave a residual even in the first bending mode.

The indices also describe the coexistence of bending- and torsion-like displacement components. On the plate, mixed displacement patterns are observed in modes 5 and 6, whose frequencies differ by 2.5~Hz. On the XB-2, the second and fourth modes have twist shares of 0.451 and 0.339 and are classified as coupled, consistent with the labels of the original benchmark \cite{Dessena2022b}.

Likewise, based on the measured components, the in-plane ratio classifies the identified mode
at 15.25~Hz, whose modal parameters were not identified in the earlier campaign, as in-plane-dominant.
For the building case study, the same two indices apply once the height is taken as the span axis. There, $\tau$ classifies the five modes as bending modes, and $\rho$ measures their East--West motion, which the original identification notes as a skew of the first mode.\cite{Dessena2024f} Although the predominantly translational Sheraton modes
do not provide a direct validation for torsional modes, this has already been assessed on the plate and XB-2 wing applications, and the investigated structure is well representative of similar multi-storey multi-bay RC frames.

These results define a rule that assigns a modal label using $\tau$ and the global and section-wise in-plane ratios. The dominant motion is in-plane bending when $\rho > 1$, out-of-plane bending when $\rho \le 1$ and $\tau < 1/2$, and torsion when $\rho \le 1$ and $\tau \geq 1/2$. 
Each label lists the dominant motion first, followed by any secondary motion introduced by ``coupled with''.
The in-plane/out-of-plane ordering follows the peak-amplitude
criterion, whereas the ordering of out-of-plane bending and
torsion follows the squared pair-displacement shares.
At $\tau=1/2$, the two displacement shares are equal and ``torsion'' is listed first solely as a deterministic tie convention.
At $\rho=1$, the branch defined by $\rho\leq1$ is used.
Out-of-plane bending and torsion are both included in the label when
the smaller of $\tau$ and $1 - \tau$ is at least a bound $\tau_b$, in the order of their shares. 
Similarly, the in-plane and out-of-plane motions are both included in the label when
the section in-plane ratio $\rho_s$, given by \cref{eq:rho} over the two sensors of section $s$, lies between a bound $\rho_b$ and $1/\rho_b$ at two or more sections. This section-wise check is needed because $\rho$ compares the largest motions only, which may occur in different sections.
This test detects local comparability between the measured in-plane and out-of-plane amplitudes.
Sections outside the prescribed band do not contribute, even when one component is locally large.
Increasing $\rho_s$ beyond $1/\rho_b$ therefore removes that section from the count, making the test non-monotonic with respect to changes in a single component.

The two bounds are calibrated using the 17 modes
of the three cases presented here.
In the modes without coupling between out-of-plane bending and torsion, the smaller of $\tau$ and $1 - \tau$ is at most 0.03, whereas in the modes with this coupling it is at least 0.10. Hence, $\tau_b$ is set at 0.05, within this gap. 
Likewise, in the modes without a secondary in-plane or out-of-plane motion, 
at most one section has $\rho_s$ above 0.38, whereas in the other modes at least two sections have $\rho_s$ from 0.85 to 2. Hence, $\rho_b$ is set at 1/2, so that $\rho_s$ is counted from half to twice the equality point. 
With these bounds, the rule assigns the labels reported for the 17 modes in \cref{tab:fe,tab:xb2,tab:sheraton}, reproducing the visual classifications of the eight plate modes.
With $\rho_b=0.5$ and the remaining criteria fixed, the tabulated modes retain their labels for $0.04\leq\tau_b\leq0.09$.

The method has three limitations, related to, respectively, geometry considerations, missing directional components, and the transferability of the so-determined thresholds. 

First, the indices require that the span axis of the structure be at least partially defined, through its direction and a reference point, since $T$, the sections of $\tau$ and the reference plane of $\rho$ are all referred to. In addition, the twist share requires a pair of sensors transversely spaced to the span axis in each section. In a layout without such pairs, the classification is based on $B$ and $T$, and on $\rho$ only when $\hat{\mathbf{n}}$ is prescribed as the normal of a known reference plane. The twist is also measured about the midpoint of each pair, which coincides with the twist axis only when the two sensors lie symmetrically about it. Since $T$ is in turn referred to the span axis, both $T$ and $\tau$ are distorted when the twist axis of the structure is not well defined or lies away from these references. On the XB-2, the pairs are not symmetric about the quarter-chord line. The shear centre of the wing, taken here as an estimate of the twist axis, lies from about 20 mm to 8 mm aft of that line at the four stations.\cite{Dessena2022h} Hence, on this layout, $\tau$ and $T$ of a pure torsion mode depart from their ideal value of 1. With the positions and the measured components in \cref{tab:xb2sensors}, a rigid rotation of all sections by the same angle about the quarter-chord line gives $B = 0.76$, $T = 0.45$, $\tau = 0.64$ and $\rho = 0.12$. The same rotation about the shear-centre line gives $B = 0.47$, $T = 0.49$, $\tau = 0.83$ and $\rho = 0.15$. Thus, $\tau$ of a torsion mode is expected at about 0.6 to 0.8, not at 1, and the twist shares of the second and fourth modes, 0.451 and 0.339, lie below this range. Hence, the equality point $\tau = 1/2$ of the rule does not account for this reduced value. 
As a sensitivity check, setting the threshold for $\tau$ to half the pure-rotation reference value gives approximately 0.32 for the quarter-chord line and 0.42 for the shear-centre line. The twist share of mode 2 ($\tau = 0.451$) exceeds both thresholds, whereas that of mode 4 ($\tau = 0.339$) exceeds only the threshold for the quarter-chord line.
For such a rotation, $T$ is close to the value of the first bending mode, 0.466, such that it does not separate torsion from bending on this layout. 

Second, incomplete directional measurements make $B$, $T$, and $\rho$ dependent on the available channels. This is relevant to the XB-2, where chordwise motion is measured at only three of the eight accelerometers. Triaxial measurements at every sensor would remove the dependence on missing components, while the dependence on sensor geometry would remain.
{The sensitivity of the classification to sensor coverage is also examined on the building, by omitting one instrumented floor at a time from the sampled modal vectors, with the identified modes and the bounds fixed. Nineteen of the 20 classifications retain their label. Omitting the roof changes mode 5 from North--South bending coupled with East--West bending to the reverse order, as $\rho$ increases from 0.27 to 1.75; this mode is the least clear-cut also with all floors.}

Third, $\tau_b$ and $\rho_b$ are derived from the three cases presented here, and may need to be revised for other structures. 

\section{Conclusions}\label{sec:conclusions}
This work introduces four indices that describe the motion of a mode from the motion vectors of a small set of sensors, without the visual inspection of the mode shape. They are a bending score, a torsion score, a twist share and an in-plane ratio, and the last two are used to classify each mode through the rule stated in Section~\ref{sec:discussion}. The main findings are:
\begin{itemize}
  \item On the first eight modes of a tapered plate with a swept trailing edge, the twist share separates the uncoupled bending modes, with $\tau$ not above 0.009, from the uncoupled torsion modes, with $\tau$ of 0.971 and 0.996. 
  It also quantifies the coupling of modes 5 to 7, which exhibit both symmetric and antisymmetric displacement components, with twist shares of 0.101, 0.900 and 0.141.
  The in-plane ratio stays below 0.02 for all the modes of the plate, which are out-of-plane;
  \item On the XB-2 wing, four modes are identified through the real-valued iLF, and the three modes with a counterpart in the original benchmark lie within 0.7\% of its frequencies. The indices classify one bending mode ($\tau =$ 0.001) and three coupled modes: two of out-of-plane bending and torsion ($\tau =$ 0.451 and 0.339), and one classified as in-plane-dominant according to the
peak-amplitude criterion ($\rho=1.33$).
  The mode dominated by in-plane motion at 15.25 Hz, not investigated in the earlier campaigns, is, to the authors' knowledge, the first in-plane mode of the XB-2 wing with identified modal parameters;
  \item On the Sheraton Universal Hotel, with the height taken as the span axis, $\tau$ stays below 0.0005 and classifies the five modes as North--South bending. In addition, $\rho$ of 0.21 to 0.53 measures their East--West motion, comparable to the North--South motion at several floors in the fourth and fifth modes;
  \item The bending score measures the coherence of the motion, with values of 1.000, 0.996 and 1.000 for the first bending modes of the plate, of the XB-2 and of the building. The torsion score depends on the position and the measured components of the sensors. Hence, each mode is classified using the rule based on the twist share and the in-plane ratio. Its bounds, $\tau_b =$ 0.05 and $\rho_b =$ 1/2, are derived from the data of the three cases.
\end{itemize}
Future work will apply the indices to triaxial measurements at every sensor and to operational data from long-term monitoring.

\section*{Data and software availability}

\subsection*{Source data}
The mode shapes of the Sheraton Universal Hotel (Data supporting: NExT-LF: A Novel Operational Modal Analysis Method via Tangential Interpolation -- Data file 1), identified from its ambient vibration tests in earlier work,\cite{Dessena2024f} are openly available from the Zenodo repository at \url{https://doi.org/10.5281/zenodo.14975925} under the terms of the [GNU GPL 3.0].

\subsection*{Underlying data}
{The XB-2 acceleration records and transfer function processed in this study (A dataset for partial in-plane acceleration of the XB-2 wing -- Data file 2) are openly available from the Zenodo repository at \url{https://doi.org/10.5281/zenodo.24536230} under the terms of the [CC BY 4.0].}

\subsection*{Software availability}
{The code used in this work (Accompanying data: Automated classification of modal displacements -- Data file 3) is openly available from the Zenodo repository at \url{https://doi.org/10.5281/zenodo.24539201} under the terms of the [MIT licence].}

\section*{AI-use declaration}
During the preparation of this work, the authors used Claude (Anthropic) Opus 5.5 to debug the TikZ code of Figures 1, 2, and 4. The authors reviewed and edited all content and take full responsibility for the content of the published article.

\section*{Competing interests}
No competing interests to disclose.

\section*{Acknowledgements}
The Department of Aerospace Engineering at the Universidad Carlos III de Madrid has been accredited as a Unit of Excellence ``Mar\'ia de Maeztu'', a distinction awarded by the Spanish State Research Agency from 1 January 2026 to 31 December 2031.

\bibliographystyle{ore}   

\begin{thebibliography}{10}

\bibitem{Ewins2000}
Ewins DJ: \textbf{\emph{Modal Testing: Theory, Practice and Application.}} 2nd ed. Research Studies Press; 2000.

\bibitem{Fan2011}
Fan W, Qiao P: \textbf{Vibration-based Damage Identification Methods: A Review and Comparative Study.} \emph{Structural Health Monitoring.} 2011; \textbf{10}(1): 83--111. \newblock \oredoilink{10.1177/1475921710365419}

\bibitem{Scalia2026}
Scalia L, Cavallaro R, Cini A: \textbf{Structural sensitivity analysis assisted by automatic differentiation.} \emph{Structural and Multidisciplinary Optimization.} 2026; \textbf{69}(7): 177. \newblock \oredoilink{10.1007/s00158-026-04365-z}

\bibitem{Dessena2025}
Dessena G, Pontillo A, Civera M, \emph{et~al.}: \textbf{Damping Identification Sensitivity in Flutter Speed Estimation.} \emph{Vibration.} 2025; \textbf{8}(2): 24. \newblock \oredoilink{10.3390/vibration8020024}

\bibitem{Modesti2024}
Modesti M, Gentilini C, Palermo A, \emph{et~al.}: \textbf{A two-step procedure for damage detection in beam structures with incomplete mode shapes.} \emph{Journal of Civil Structural Health Monitoring.} 2025; \textbf{15}(2): 287--306. \newblock \oredoilink{10.1007/s13349-024-00839-0}

\bibitem{Vila2024}
Vila JLCR, Carneiro SHS, Goulart JNV, \emph{et~al.}: \textbf{Wavelet-based numerical investigation on damage localisation and quantification in beams using static deflections and mode shapes.} \emph{European Journal of Mechanics - A/Solids.} 2024; \textbf{107}: 105351. \newblock \oredoilink{10.1016/j.euromechsol.2024.105351}

\bibitem{Georgioudakis2018}
Georgioudakis M, Plevris V: \textbf{A Combined Modal Correlation Criterion for Structural Damage Identification with Noisy Modal Data.} \emph{Advances in Civil Engineering.} 2018; \textbf{2018}: 3183067. \newblock \oredoilink{10.1155/2018/3183067}

\bibitem{Dessena2025a}
Dessena G, Civera M, Marcos A, \emph{et~al.}: \textbf{Multiple input tangential interpolation-driven damage detection of a jet trainer aircraft.} \emph{Aerospace Science and Technology.} 2026; \textbf{168}: 111032. \newblock \oredoilink{10.1016/j.ast.2025.111032}

\bibitem{BukaVaivade2025}
Buka-Vaivade K, Nicoletti V, Gara F: \textbf{Advancing bridge resilience: a review of monitoring technologies for flood-prone infrastructure.} \emph{Open Research Europe.} 2025; \textbf{5}: 26. Version 2. \newblock \oredoilink{10.12688/openreseurope.19232.2}

\bibitem{Tantaroudas2026}
Tantaroudas ND, Zamora-Sanchez D, Armijo JA, \emph{et~al.}: \textbf{A crowdsensing platform for structural health monitoring of rural bridge infrastructure.} \emph{Open Research Europe.} 2026; \textbf{6}: 50. \newblock \oredoilink{10.12688/openreseurope.22915.2}

\bibitem{Mottershead1993}
Mottershead JE, Friswell MI: \textbf{Model Updating In Structural Dynamics: A Survey.} \emph{Journal of Sound and Vibration.} 1993; \textbf{167}(2): 347--375. \newblock \oredoilink{10.1006/jsvi.1993.1340}

\bibitem{Goge2003}
Göge D, Link M: \textbf{Assessment of computational model updating procedures with regard to model validation.} \emph{Aerospace Science and Technology.} 2003; \textbf{7}(1): 47--61. \newblock \oredoilink{10.1016/S1270-9638(02)01193-8}

\bibitem{Ewins2000a}
Ewins DJ: \textbf{Model validation: correlation for updating.} \emph{Sadhana.} 2000; \textbf{25}(3): 221--234. \newblock \oredoilink{10.1007/BF02703541}

\bibitem{Allemang1982}
Allemang RJ, Brown DL: \textbf{A correlation coefficient for modal vector analysis.} In: \emph{Proceedings of the 1st International Modal Analysis Conference.} Union College; 1982. pp. 110--116. \newblock \oreurllink{https://web.archive.org/web/20151018074821/http://sdrl.uc.edu/sdrl/referenceinfo/documents/papers/imac1982-mac.pdf}

\bibitem{Lieven1988}
Lieven NAJ, Ewins DJ: \textbf{Spatial Correlation of Mode Shapes, the Coordinate Modal Assurance Criterion (COMAC).} In: \emph{Proceedings of the 6th International Modal Analysis Conference.} Society for Experimental Mechanics; 1988. pp. 690--695.

\bibitem{Verboven2002}
Verboven P, Parloo E, Guillaume P, \emph{et~al.}: \textbf{Autonomous structural health monitoring - Part I: Modal parameter estimation and tracking.} \emph{Mechanical Systems and Signal Processing.} 2002; \textbf{16}(4): 637--657. \newblock \oredoilink{10.1006/mssp.2002.1492}

\bibitem{Volkmar2026}
Volkmar R, Soal K, Baum M, \emph{et~al.}: \textbf{Robust online monitoring of aircraft modal parameters using data fusion-based mode tracking.} \emph{Mechanical Systems and Signal Processing.} 2026; \textbf{250}: 114171. \newblock \oredoilink{10.1016/j.ymssp.2026.114171}

\bibitem{Wright2014}
Wright JR, Cooper JE: \textbf{\emph{Introduction to Aircraft Aeroelasticity and Loads.}} Wiley; 2014. \newblock \oredoilink{10.1002/9781118700440}

\bibitem{Goge2003a}
Göge D: \textbf{Automatic updating of large aircraft models using experimental data from ground vibration testing.} \emph{Aerospace Science and Technology.} 2003; \textbf{7}(1): 33--45. \newblock \oredoilink{10.1016/S1270-9638(02)01184-7}

\bibitem{Kerschen2013}
Kerschen G, Peeters M, Golinval JC, \emph{et~al.}: \textbf{Nonlinear modal analysis of a full-scale aircraft.} \emph{Journal of Aircraft.} 2013; \textbf{50}(5): 1409--1419. \newblock \oredoilink{10.2514/1.C031918}

\bibitem{Das2026}
Das A, Yao J, Coppotelli G, \emph{et~al.}: \textbf{Ground Vibration Test Analysis of a PC-9 Aircraft using Delay-embedded Dynamic Mode Decomposition with Control.} In: \emph{35th Congress of the International Council of the Aeronautical Sciences, ICAS 2026.} International Council of the Aeronautical Sciences; 2026. \newblock \oredoilink{10.71945/icas2026_0153}

\bibitem{Pontillo2025}
Pontillo A, Healy F, Gu H, \emph{et~al.}: \textbf{Experimental Testing and Analysis of a Very Flexible Wing Model.} In: \emph{AIAA SCITECH 2025 Forum.} American Institute of Aeronautics and Astronautics; 2025. pp. 1--19. \newblock \oredoilink{10.2514/6.2025-0424}

\bibitem{Hancock1985}
Hancock GJ, Wright JR, Simpson A: \textbf{On the teaching of the principles of wing flexure-torsion flutter.} \emph{The Aeronautical Journal.} 1985; \textbf{89}(888): 285--305. \newblock \oredoilink{10.1017/S0001924000015050}

\bibitem{Dessena2022b}
Dessena G, Ignatyev DI, Whidborne JF, \emph{et~al.}: \textbf{Ground vibration testing of a flexible wing: A benchmark and case study.} \emph{Aerospace.} 2022; \textbf{9}(8): 438. \newblock \oredoilink{10.3390/aerospace9080438}

\bibitem{Dessena2022h}
Dessena G, Pontillo A, Ignatyev DI, \emph{et~al.}: \textbf{Identification of Nonlinearity Sources in a Flexible Wing.} \emph{Journal of Aerospace Engineering.} 2025; \textbf{38}(5): 04025060. \newblock \oredoilink{10.1061/JAEEEZ.ASENG-5508}

\bibitem{Gueguen2021}
Gu{\'e}guen P, Guattari F, Aubert C, \emph{et~al.}: \textbf{Comparing Direct Observation of Torsion with Array-Derived Rotation in Civil Engineering Structures.} \emph{Sensors.} 2021; \textbf{21}(1): 142. \newblock \oredoilink{10.3390/s21010142}

\bibitem{Aulakh2021}
Aulakh DS, Bhalla S: \textbf{{3D} torsional experimental strain modal analysis for structural health monitoring using piezoelectric sensors.} \emph{Measurement.} 2021; \textbf{180}: 109476. \newblock \oredoilink{10.1016/j.measurement.2021.109476}

\bibitem{Pappa1993}
Pappa RS, Elliott KB, Schenk A: \textbf{Consistent-mode indicator for the eigensystem realization algorithm.} \emph{Journal of Guidance, Control, and Dynamics.} 1993; \textbf{16}(5): 852--858. \newblock \oredoilink{10.2514/3.21092}

\bibitem{Reynders2012a}
Reynders E, Houbrechts J, {De Roeck} G: \textbf{Fully automated (operational) modal analysis.} \emph{Mechanical Systems and Signal Processing.} 2012; \textbf{29}: 228--250. \newblock \oredoilink{10.1016/j.ymssp.2012.01.007}

\bibitem{Coppolino2011}
Coppolino RN: \textbf{Understanding Large Order Finite Element Model Dynamic Characteristics.} In: Proulx T (Ed.), \emph{Linking Models and Experiments, Volume 2.} New York, NY: Springer; 2011. pp. 367--373. \newblock \oredoilink{10.1007/978-1-4419-9305-2_26}

\bibitem{Larsen2002}
Larsen GC, Hansen MH, Baumgart A, \emph{et~al.}: \textbf{Modal analysis of wind turbine blades.} Technical Report Ris{\o}-R-1181(EN), Ris{\o} National Laboratory. 2002. \newblock \oreurllink{https://orbit.dtu.dk/en/publications/modal-analysis-of-wind-turbine-blades-2/}

\bibitem{Chen2022}
Chen Y, Griffith DT: \textbf{Finite Cross-Section Method for Mode Shape Recognition of Highly Coupled Beam-Type Structures.} \emph{Journal of Vibration and Acoustics.} 2022; \textbf{144}(4): 041013. \newblock \oredoilink{10.1115/1.4053977}

\bibitem{Tohmuang2025}
Tohmuang S, Fard M, Marzocca P, \emph{et~al.}: \textbf{Structure mode shapes classification using graph convolutional networks in automotive application.} \emph{Computers \& Structures.} 2025; \textbf{314}: 107767. \newblock \oredoilink{10.1016/j.compstruc.2025.107767}

\bibitem{Dessena2024}
Dessena G, Civera M: \textbf{Improved tangential interpolation-based multi-input multi-output modal analysis of a full aircraft.} \emph{European Journal of Mechanics - A/Solids.} 2025; \textbf{110}: 105495. \newblock \oredoilink{10.1016/j.euromechsol.2024.105495}

\bibitem{Yusuf2019}
Yusuf S, Pontillo A, Weber S, \emph{et~al.}: \textbf{Aeroelastic Scaling for Flexible High Aspect Ratio Wings.} In: \emph{AIAA Scitech 2019 Forum.} San Diego, CA: American Institute of Aeronautics and Astronautics; 2019. \newblock \oredoilink{10.2514/6.2019-1594}

\bibitem{Abbott1945}
Abbott IH, von Doenhoff AE, Stivers Jr. LS: \textbf{Summary of Airfoil Data.} NACA Report 824, National Advisory Committee for Aeronautics. 1945. NASA NTRS 19930090976. \newblock \oreurllink{https://ntrs.nasa.gov/citations/19930090976}

\bibitem{Dessena2024f}
Dessena G, Civera M, Yousefi A, \emph{et~al.}: \textbf{{NExT-LF}: A Novel Operational Modal Analysis Method via Tangential Interpolation.} \emph{International Journal of Mechanical System Dynamics.} 2025; \textbf{5}(3): 401--414. \newblock \oredoilink{10.1002/msd2.70016}

\end{thebibliography}
\providecommand{\oredoilink}[1]{\href{https://doi.org/#1}{Publisher Full Text}}
\providecommand{\oreurllink}[1]{\href{#1}{Reference Source}}

\end{document}